\documentclass[twocolumn,nosuperscriptaddress,nofootinbib,10pt]{revtex4-1}
\usepackage[perpage]{footmisc} 
\usepackage{epstopdf}
\usepackage[dvipsnames]{xcolor}
\usepackage{epsfig}
\usepackage{color}
\usepackage{hyperref}
\usepackage{mathbbol}
\usepackage{bookmark}
\usepackage{cjhebrew}
\usepackage{amsmath}
\usepackage{amssymb}
\usepackage{slashed}
\usepackage{nicefrac}
\usepackage{tikz,pgfplots}
\usepgfplotslibrary{patchplots}
\pgfplotsset{compat=1.10}
\pgfplotsset{
  cuslegR/.style={legend image code/.code={
\draw[red,dash pattern=on 1.5pt off 1.5pt on 1.5pt off 1.5pt] (0cm,0cm)     -- (0.5cm,0cm);
\draw[red,solid]  (0.5cm, 0cm) -- (1.cm, 0cm);
}}}
\pgfplotsset{
  cuslegG/.style={legend image code/.code={
\draw[green,dash pattern=on 1.5pt off 1.5pt on 1.5pt off 1.5pt] (0cm,0cm)     -- (0.5cm,0cm);
\draw[green,solid]  (0.5cm, 0cm) -- (1.cm, 0cm);
}}}
\pgfplotsset{
  cuslegB/.style={legend image code/.code={
\draw[blue,dash pattern=on 1.5pt off 1.5pt on 1.5pt off 1.5pt] (0cm,0cm)     -- (0.5cm,0cm);
\draw[blue,solid]  (0.5cm, 0cm) -- (1.cm, 0cm);
}}}
  \usepgfplotslibrary{fillbetween}
\usetikzlibrary{positioning}
\usetikzlibrary{calc}
\usetikzlibrary{arrows,shapes,backgrounds}
\pgfdeclarepatternformonly{north east lines wide}%
   {\pgfqpoint{-1pt}{-1pt}}%
   {\pgfqpoint{10pt}{10pt}}%
   {\pgfqpoint{9pt}{9pt}}%
   {
     \pgfsetlinewidth{1.4pt}
     \pgfpathmoveto{\pgfqpoint{0pt}{0pt}}
     \pgfpathlineto{\pgfqpoint{9.1pt}{9.1pt}}
     \pgfusepath{stroke}
    }
\tikzset{
  pics/carc/.style args={#1:#2:#3}{
    code={
      \draw[pic actions] (#1:#3) arc(#1:#2:#3);
    }
  }
}
\usetikzlibrary{decorations.pathreplacing}
\usetikzlibrary{decorations.pathmorphing,patterns}
\usetikzlibrary{decorations.markings,calc}
\usetikzlibrary{shapes.callouts}
\tikzset{
    level/.style = {
        thick,
    },
    connect/.style = {
        dotted,
    },
    notice/.style = {
        draw,
        rectangle callout,
        callout relative pointer={#1}
    },
    label/.style = {
        text width=2cm
    }
}
\pgfdeclaredecoration{discontinuity}{start}{
  \state{start}[width=0.75\pgfdecoratedinputsegmentremainingdistance-0.5\pgfdecorationsegmentlength,next state=first wave]
  {}
  \state{first wave}[width=\pgfdecorationsegmentlength, next state=second wave]
  {
    \pgfpathlineto{\pgfpointorigin}
    \pgfpathmoveto{\pgfqpoint{0pt}{\pgfdecorationsegmentamplitude}}
    \pgfpathcurveto
        {\pgfpoint{-0.25*\pgfmetadecorationsegmentlength}{0.75\pgfdecorationsegmentamplitude}}
        {\pgfpoint{-0.25*\pgfmetadecorationsegmentlength}{0.25\pgfdecorationsegmentamplitude}}
        {\pgfpoint{0pt}{0pt}}
    \pgfpathcurveto
        {\pgfpoint{0.25*\pgfmetadecorationsegmentlength}{-0.25\pgfdecorationsegmentamplitude}}
        {\pgfpoint{0.25*\pgfmetadecorationsegmentlength}{-0.75\pgfdecorationsegmentamplitude}}
        {\pgfpoint{0pt}{-\pgfdecorationsegmentamplitude}}
}
\state{second wave}[width=0pt, next state=do nothing]
  {
    \pgfpathmoveto{\pgfqpoint{0pt}{\pgfdecorationsegmentamplitude}}
    \pgfpathcurveto
        {\pgfpoint{-0.25*\pgfmetadecorationsegmentlength}{0.75\pgfdecorationsegmentamplitude}}
        {\pgfpoint{-0.25*\pgfmetadecorationsegmentlength}{0.25\pgfdecorationsegmentamplitude}}
        {\pgfpoint{0pt}{0pt}}
    \pgfpathcurveto
        {\pgfpoint{0.25*\pgfmetadecorationsegmentlength}{-0.25\pgfdecorationsegmentamplitude}}
        {\pgfpoint{0.25*\pgfmetadecorationsegmentlength}{-0.75\pgfdecorationsegmentamplitude}}
        {\pgfpoint{0pt}{-\pgfdecorationsegmentamplitude}}
    \pgfpathmoveto{\pgfpointorigin}
}
  \state{do nothing}[width=\pgfdecorationsegmentlength,next state=do nothing]{
    \pgfpathlineto{\pgfpointdecoratedinputsegmentlast}
  }
  \state{final}
  {
    \pgfpathlineto{\pgfpointdecoratedpathlast}
  }
}
\definecolor{blue}{HTML}{4169E1}
\definecolor{red}{HTML}{DC143C}
\definecolor{green}{HTML}{2E8B57}
\definecolor{black}{HTML}{000000}
\definecolor{g1}{HTML}{A9A9A9}
\definecolor{g2}{HTML}{696969}
\definecolor{g3}{HTML}{7F7F7F}
\definecolor{g4}{HTML}{D3D3D3}

\newcommand{\eps}{\epsilon}

\newcommand{\at}{a_t}
\newcommand{\ecm}{E_\textrm{\small c.m.}}

\newcommand{\mpil}{$m_\pi\approx806~${\small MeV}}

\newcommand{\eg}{\textit{e.g.}\;}
\newcommand{\ie}{\textit{i.e.}\;}
\newcommand{\be}{\begin{equation}}
\newcommand{\ee}{\end{equation}}
\newcommand{\la}{\label}
\newcommand{\ber}{\begin{eqnarray}}
\newcommand{\eer}{\end{eqnarray}}

\newcommand{\bea}{\begin{eqnarray}}
\newcommand{\eea}{\end{eqnarray}}
\newcommand{\beq}{\begin{align}}
\newcommand{\eeq}{\end{align}}

\newcommand{\bt}{B_{^{3}\text{H}}}

\newcommand{\bd}{B_\text{D}}

\newcommand{\lam}[1]{$\Lambda=#1~$fm$^{-1}$}

\newcommand{\ve}[1]{\ensuremath{\boldsymbol{#1}}}

\newcommand{\eftnopi}{\mbox{EFT($\slashed{\pi}$)}}

\newcommand{\parag}[1]{\paragraph*{\bf\textit{#1}:}}
\DefineFNsymbols*{lamportnostar}[math]{\dagger\ddagger\S\P\|{\dagger\dagger}{\ddagger\ddagger}}
\setfnsymbol{lamportnostar}

\begin{document}
\title{Emergence of a Brunnian neutron state}
%
\author{Johannes~Kirscher}
\email{jkirscher@ccny.cuny.edu}
\affiliation{Department of Physics, The City College of New York, New York, NY 10031, USA}
\date{\today}
\begin{abstract}
We discuss a quantum-statistical feature of non-relativistic
identical fermions whose interaction is predominantly attractive at low
energies.
Specifically, we consider exotic, multi-neutron nuclei.
From the enhancement of an arbitrarily small $P$-wave
interaction between two nucleons, we infer the existence of
\emph{a} particle-stable nucleus
composed entirely of neutrons. While we cannot
specify the number of neutrons in the system, we
predict that none of its substructures is bound.
The independence of this deduction from the
short-distance structure of the nuclear interaction and its consistency
with deuteron, triton, and helium-4 properties is established. 
\end{abstract}
\maketitle
$${}$$
\parag{Overture}\la{sec:overt}
The inter-nuclear interaction is not known accurately enough
to make reliable predictions about the existence of stable, multi-neutron
clusters. An improved understanding of such exotic structures transcends
to equally intriguing systems of contemporary theoretical and experimental
interest, \eg, neutron stars and neutron-rich nuclei near and beyond the
drip line of stability.
With the major impediment for advancement in this sector being the
extreme difficulties in obtaining data of similar accuracy
for the uncharged and unstable neutrons as for nuclei which incorporate
protons, it were, unsurprisingly, recent
\emph{experiments}~\cite{PhysRevLett.116.052501,BYSTRITSKY2016164} which reinvigorated research\footnote{See
Refs.~\cite{PhysRevLett.90.252501,PhysRevC.72.034003,Carbonell2017,
Fossez:2016dch,Gandolfi:2016bth} and the exemplary
studies~\cite{PhysRevC.60.024002,PhysRevC.66.054001} which preceded the recent
experiments.}
in this field.
Prior to these modern approaches which draw from the decade-long development of
high-precision nuclear interaction models and effective field theories,
universal mechanisms and phenomena related to multi neutrons were
envisioned by A.~Migdal~\cite{Migdal:1972ss}. In particular,
he considered the formation of stable di-neutrons (${}^2n$)
in the force field of a lithium-9 core.

Here, we replace this stabilizing nuclear core with a set of neutrons which
is unstable in isolation.
We show that it can be expected to be glued together with orbiting neutrons,
and as such, constitutes a neutron analog of the molecular hydrogen ion.
Naturally, we are also constrained by the uncertainty in two-neutron 
data, but employing an effective field theory (EFT)
allows us to assess the sensitivity of the neutron states
with respect to precisely such modifications of the nuclear interaction
which reflect this ignorance while still being
consistent in the description of small nuclei for which data is accurate.
Furthermore, the EFT framework allows for a generalization of the result
to a class of equal-mass fermionic systems with an attractive short-range
potential of insufficient strength to form a bound dimer, and hence the
main conjecture of the article, namely, the emergence of a bound state without
bound subsystems, can be tested with non-nuclear particles.

Pertinent to \emph{this} work,
W.~Heisenberg's original approximation of
an isospin-independent nuclear force is sufficient.
All differences between multi-neutron states
and ordinary nuclei can in this approximation be attributed to
quantum statistics: the demand of a totally anti-symmetric wave function
reduces the dimension of the state space available to multiple neutrons
severely compared to a mixture of protons and neutrons. The two-nucleon (NN) configuration
of strongest nuclear attraction, supporting the deuteron, is thereby forbidden
for two neutrons. The formation of larger neutron states by an agglomeration
on a bound seed, like the deuteron, is thus inhibited.
There is, however, precedent for bound systems which
contain only unbound substructures.
In the hydrogen molecule, a single electron combines with a system
of otherwise unbound protons. Even more apt is the helium-6 nucleus, which can
be understood as a three-body system with a helium-4 core plus two neutrons. While
neither helium-5 nor the two neutrons form bound states\footnote{In
general, a bound $N$-body structure without any stable sub-configuration
is named to honor its original treatment by H.~Brunn~\cite{brunni}.
If $N=3$, the state is called Borromean. Ref.~\cite{Baas:2012za}~discusses
various physical realizations of such structures.},
the compound does.
It is this note,  we demonstrate that neutrons
behave similarly in the sense that there exists a critical number
(\cjRL{b*}$-1$)~of neutrons which are bound together by
the addition of another neutron.
The accuracy of our study does not allow for a prediction of \cjRL{b*}~but
the logical conclusion guarantees
\cjRL{b*}~$<\infty$ and thereby the existence of a bound, pure neutron
state\footnote{At the scale of
strong interaction processes, \ie, we do not consider electro-weak effects.}.

To that end, we specify the interaction between two neutrons, first. Subsequently,
we test for its analogy to the Coulomb interaction in the hydrogen
molecule, namely, whether it bears the potential to affect
binding of larger systems despite of its failure to bind two, three,
four, up to~\cjRL{b*}$-1$ neutrons.
We confirm this molecular character~\cite{Wheeler:1937zza}~of the nuclear interaction by
quantifying the enhancement of an arbitrarily small disturbance of the
NN potential in the three and four-neutron systems.
For the tri-neutron, the $J^\pi=\nicefrac{1}{2}^-$ configuration is identified
as the channel most amenable to experimental detection and $0^+$ for the
tetra-neutron.
From the invariance of the enhancement of the $P$-wave interaction
with respect to a
renormalization-group transformation, and under the assumption that
the increase in attraction is correlated with the number of interacting pairs in
the system, we infer the Brunnian character of neutrons.

\parag{Interaction}\la{sec:int}
To describe the interaction between neutrons, we invoke a minimal
theory that provides reliable uncertainty estimates for its
predictions\footnote{For details on this pion-less effective field theory (\eftnopi) see
Refs.~\cite{Kaplan:1998tg,vanKolck:1998bw,Chen:1999tn}}.
With this theory, we can guarantee consistency of our results
for few-neutron systems with all other observables it is
appropriately used for.

The Hamiltonian comprises iso-spin and momentum independent contact terms with
\be\la{eq:ham}
  \hat{H} = \sum_{i}^A \frac{-\ve{\nabla}_i^2}{2m}
                        + \sum_{i<j}^A
                       \left[c^\Lambda_S\,\hat{P}^{ij}_S+
                       c^\Lambda_T\,\hat{P}^{ij}_T
                        \right]\;\;.
\ee
In this form\footnote{$\Lambda$ parameterizes the regularization,
the operator $\hat{P}_{S(T)}$ projects onto spin-singlet(triplet) antisymmetric
NN states. For details see, \eg, Ref.~\cite{Kirscher:2015yda}, and
the supplemental material to this article.},
the theory has been successful in its description of
ground-state properties of nuclei with up to four constituents.
To our knowledge, the consideration of shallow few-neutron states
is the first application of \eftnopi~to nuclear properties of a
\emph{molecular} character.
Previous studies highlighted the sensitivity of multi-neutron states to
details of the nuclear interaction, and hence the systematic uncertainty
analysis of the EFT framework is the most important reason for its usage here.
We parametrize part of this uncertainty with a spin-orbit interaction: 
\be\la{eq:LS}
\hat{V}_\text{\tiny ls}=\eps\,c_T^\Lambda\,\ve{L}\cdot\ve{S}\;\;,
\ee
another part with the cutoff regulator $\Lambda$.
As we anticipate a molecular binding mechanism for neutrons,
this choice suggests itself given the prominent role of the term
in refinements of atomic and nuclear shell models.
Specifically, we vary $\eps$ such that the theory still predicts those
nuclear processes for which it devised with the required accuracy, namely, the
deuteron, triton, helium-3, and $\alpha$-particle binding energies,
as well as NN $P$-wave phase shifts (inset of Fig.~\ref{fig:3n-phases}).
The effect on the latter is shown in Fig.~\ref{fig:2n-phases}, where
we compare the EFT phases (dashed) for an $\eps$~large enough to produce an
unphysical, deeply-bound tetra-neutron with neutron-proton data (solid).
The EFT-predicted phases are small relative to data for $\eps=0$ (not shown).
The enhancement in the $J=0$ channel (blue dashed) induced by a non-zero $\eps$ is dominant,
and the theory clearly does not reproduce the attractive/repulsive character of the
other channels well. However, data and the $\eps$-enhanced EFT
yield similar results for the sum of the three channels,
whose smallness traditionally explains the insignificance of the $P$-waves in
non-exotic nuclei. In contrast to Refs.~\cite{PhysRevC.71.044004,PhysRevC.72.034003,PhysRevC.93.044004}, where all $P$-waves were enhanced democratically and bound
multi-neutrons emerged only after di-neutrons were bound, the selective enhancement
induced here by the short-range spin-orbit interaction binds tetra- and tri-neutrons
before di-neutrons.
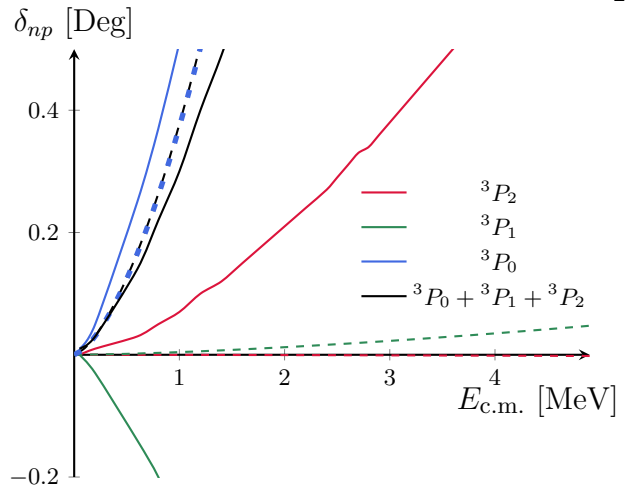
\begin{figure}
\begin{tikzpicture}[scale=1.]
    \begin{axis}
    [thick,
     smooth,
     axis x line=middle,
     axis y line=middle,     
     ymin=-0.2,ymax=.5,
     no markers,
     xlabel={\large $\ecm$ [MeV]},
     x label style={at={(axis description cs:.9,.12)},rotate=0,anchor=south},
     ylabel={\large $\delta_{np}$ [Deg]},
     y label style={at={(axis description cs:0.,1.12)},rotate=0,anchor=north},
     legend entries={{$\Lambda=2~\text{fm}^{-1}$},,,,,,{$\Lambda=15~\text{fm}^{-1}$},,,{$\Lambda=4~\text{fm}^{-1}$}},
     legend style={draw=none,at={(1.02,0.72)}},
     ]
\addplot[black,dash pattern=on 6pt off 3pt,forget plot,thick] table[x expr=\thisrowno{0},y expr=\thisrowno{1}+\thisrowno{2}+\thisrowno{3}] {PHAOUT_EFT_6_p};
\addplot[red] table[x expr=\thisrowno{0}*0.5,y expr=\thisrowno{1}] {gwu_P-phases};
\addplot[red,dashed,mark=*,forget plot] table[x expr=\thisrowno{0},y expr=\thisrowno{3}] {PHAOUT_EFT_6_p};
\addlegendentry{${}^3P_2$}
\addplot[green] table[x expr=\thisrowno{0}*0.5,y expr=\thisrowno{2}] {gwu_P-phases};
\addplot[green,dashed,mark=*,forget plot] table[x expr=\thisrowno{0},y expr=\thisrowno{2}] {PHAOUT_EFT_6_p};
\addlegendentry{${}^3P_1$}
\addplot[blue] table[x expr=\thisrowno{0}*0.5,y expr=\thisrowno{3}] {gwu_P-phases};
\addplot[blue,dash pattern=on 3pt off 6pt,mark=*,forget plot,line width=2pt] table[x expr=\thisrowno{0},y expr=\thisrowno{1}] {PHAOUT_EFT_6_p};
\addlegendentry{${}^3P_0$}
\addplot[black,thick] table[x expr=\thisrowno{0}*0.5,y expr=\thisrowno{1}+\thisrowno{2}+\thisrowno{3}] {gwu_P-phases};
\addlegendentry{${}^3P_0+{}^3P_1+{}^3P_2$}
\end{axis}
\end{tikzpicture}
\caption{\small Energy dependence of the neutron-proton $P$-wave phase shifts with \lam{6}.
The GWU analysis (solid lines) is compared with \eftnopi~(dashed) at $\eps\approx\eps({}^4n)$.
The sum of phases over the three channels is shown in black for the respective model.
\la{fig:2n-phases}}
\end{figure}
\parag{Effective interaction(s)}\la{sec:eff-int}
In order to assess whether this particular interaction
exhibits molecular dynamics of neutrons, we appeal to a non-canonical variant of the
Born-Oppenheimer method. Instead of fixing the relative distance
$\ve{R}$ between the unbound two neutrons, considering the third's
motion in this two-center field, and subsequently treating $\ve{R}$ as
a variational parameter, we exploit a fixation as induced by
detuning the fundamental quark masses which is accessible quantitatively
from lattice quantum chromodynamics (QCD)~\cite{Beane:2012vq,Yamazaki:2012hi}.
The quark masses used in Ref.~\cite{Beane:2012vq} imply
unrealistically heavy nucleons and pions whose interaction supports a bound
di-neutron with $B(^2n,{}^1S_0)\approx16$~MeV. This allows
for a study of the effective interaction between three neutrons in elastic
neutron-di-neutron scattering. Qualitatively, we find a result
analogous to the bombardment of $\alpha$ particles with
neutrons.
The interaction is repulsive in the $S$-wave (negative phase shifts
in Fig.~\ref{fig:3n-phases}) and attractive in the $P$-wave
(hatched band) of the relative motion between projectile and
target -- a result one expects, because both targets,
$\alpha$ and ${}^2n$,
reside in closed shells which cannot be penetrated with the considered
projectile energies while an asymmetric approach of the neutron
in coordinate space is Pauli allowed\footnote{We find our results with a neutron target also
consistent with the helion/triton reactions in Ref.~\cite{PhysRevC.79.044606}.}.

\begin{figure}
\begin{tikzpicture}
    \begin{axis}
    [name=mainplot,
     thick,
     smooth,
      axis x line=middle,
      axis y line=middle,
     no markers,
     xmin=-0.05,xmax=5.8,
     xlabel={\large $\ecm$ [MeV]},
     x label style={at={(axis description cs:.9,.35)},rotate=0,anchor=south},
     ylabel={\large $\delta(J^\pi)$ [Deg]},
     y label style={at={(axis description cs:-0.05,.5)},rotate=90,anchor=south},
     legend entries={},
     legend style={draw=none,fill=none,at={(.45,0.99)}},
     ]
%
%
\addplot[forget plot,blue,thick,y filter/.expression={\thisrowno{1}>0.0 ? abs(\thisrowno{1}) : (\thisrowno{1}+180) },] table[x expr=\thisrowno{0},y expr=\thisrowno{1}] {PHAOUT_3n-n_6p};
\addplot+[name path=D,gray,forget plot] table[x expr=\thisrowno{0},y expr=\thisrowno{1}]
 {PHAOUT_2n-n_6m};
\addplot+[name path=B,gray,forget plot] table[x expr=\thisrowno{0},y expr=\thisrowno{1}]
 {PHAOUT_2n-n_12m};
\addplot[lightgray,pattern=north east lines] fill between[of=B and D];
\addplot+[name path=C,gray,forget plot] table[x expr=\thisrowno{0},y expr=\thisrowno{1}]
 {PHAOUT_2n-n_6};
\addplot+[name path=A,gray,forget plot] table[x expr=\thisrowno{0},y expr=\thisrowno{1}]
 {PHAOUT_2n-n_12};
\addplot[lightgray] fill between[of=C and A];
\draw[->] (axis cs:3,-31) to[bend right=10] node[rotate=70,yshift=7] {$\Lambda\to\infty$}(axis cs:3.5,-5);s
\coordinate (2n) at (axis cs:1.,36);
\coordinate (4n) at (axis cs:5.3,38);
\node at (4n) [rotate=0] {{${}^3n{\text{\tiny P}\over}n$}};
\coordinate (3nm) at (axis cs:5.3,13);
\node at (3nm) [rotate=0] {{${}^2n{\text{\tiny P}\over}n$}};
\coordinate (3np) at (axis cs:5.3,-39);
\node at (3np) [rotate=0] {{${}^2n{\text{\tiny S}\over}n$}};
\coordinate (nim) at (axis cs:2.96,39);
\node at (nim) [rotate=0] {\textcolor{red}{GWU data}};
\coordinate (e1) at (axis cs:2.45,20);
\node at (e1) [rotate=0] {\footnotesize $\eps=0$};
\coordinate (e2) at (axis cs:2.55,30);
\node at (e2) [rotate=0] {\footnotesize $\eps=1.5$};
\coordinate (insetPosition) at (rel axis cs:0.05,0.99);
    \end{axis}
\path (mainplot.outer north west); \pgfgetlastxy{\xw}{\yw}
\path (mainplot.outer north east); \pgfgetlastxy{\xe}{\ye}
\pgfmathsetmacro\miniplotwidth{veclen(\xe-\xw,0)*0.45}    
    \begin{axis}[
     at={(insetPosition)},
     anchor={north west},
     width=\miniplotwidth,
     thick,
     smooth,
     no markers,
     xmin=-0.05,xmax=5.,
     xtick={11},
     xticklabels={10},
     ytick={3.5},
     yticklabels={$3.5^\circ$},
     legend entries={},
     legend style={draw=none,fill=none,at={(.45,0.99)}},
     y tick label style={anchor=west,color=black,xshift=1.1},
     ]
    ]
\addplot[forget plot,red,dashed,thick]
 table[x expr=\thisrowno{0}*0.5,y expr=\thisrowno{1}] {nim_phases_3P0};
\addplot+[name path=B,gray,forget plot]
 table[x expr=\thisrowno{0},y expr=\thisrowno{1}] {PHAOUT_2n_6m_ls0};
\addplot+[name path=C,gray,forget plot]
 table[x expr=\thisrowno{0},y expr=\thisrowno{1}] {PHAOUT_2n_6m_ls15};
\addplot[lightgray,pattern=dots] fill between[of=B and C];
    \end{axis}
\end{tikzpicture}
\caption{\small Energy dependence of phase shifts for elastic scattering
in the ${}^2n-n$~$\nicefrac{1}{2}^-$ (hatched band) and $\nicefrac{1}{2}^+$ (gray band)
channels with $B(^2n,0^+)\approx16~$MeV),
and the ${}^3n-n$~$0^+$ channel (solid blue) with $B(^3n,\nicefrac{1}{2}^-)\approx32~$MeV
using a contact interaction (heavy pions).
Band widths represent cutoff variations $\Lambda\in[6,12]~\text{fm}^{-1}$.
In the inset, the physical NN ${}^3P_0$ phases for $\eps\in[0,1.5]$ and \lam{6} (dot-filled area)
are compared with data (red dashed).
Over an energy range identical with the main plot, phases reach a maximum of $3.6^\circ$.
\la{fig:3n-phases}}
\end{figure}
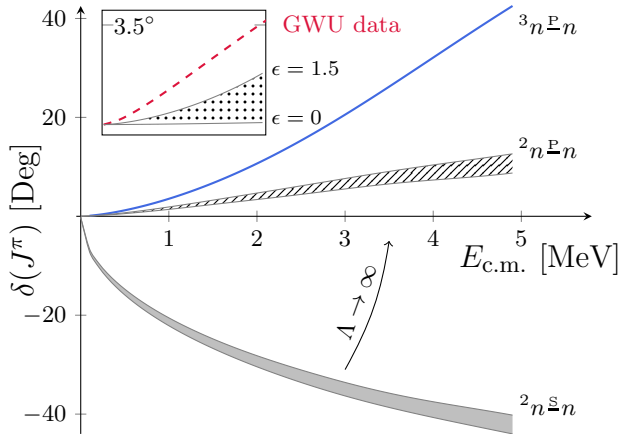

Although, the ${}^2n-n$ attraction is neither
strong enough to bind ${}^3n$ nor indicative of the presence of a shallow resonance,
it is significant in comparison with the attraction between
two neutrons, where the latter is weak relative to lattice QCD predictions~\cite{Berkowitz2017285}.
If we evoke $\hat{V}_\text{\tiny ls}$ from Eq.\eqref{eq:LS} to narrow this gap,
\ie, increase $\eps$ such that
the two-neutron phases approach the QCD values\footnote{In Fig.~\ref{fig:3n-phases}, this procedure is shown in an inset for physical two-neutron
scattering, only.},
a bound ${}^3n$ emerges, while the deuteron, the triton and helion,
and the $\alpha$ are unaffected by this variation.
We use this ${}^3n$ state as a target for a fourth neutron to
test for an enhancement of the two-neutron interaction acting
in three or four-neutron systems.

Indeed, under similar conditions, \ie, a tri-neutron bound by about 16~MeV,
the effective interaction of the neutron projectile with the ${}^3n$
is visibly stronger than with a ${}^2n$ target (compare solid blue line to hatched band in
Fig.~\ref{fig:3n-phases}). Under such conditions, a
bound ${}^4n$ is present in the spectrum, with a large enough energy not to affect
the elastic scattering process. Yet, its existence is crucial to notice because
if $\eps$ is decreased below the critical value $\eps({}^3n)$, where no
bound tri-neutron exists, this tetra-neutron remains stable. It disintegrates
into four free neutrons at some $\eps({}^4n)\ll\eps({}^3n)$.

In essence, the above shows that the effective $P$-wave interaction
between a single neutron and $(\i)$ another neutron, $(\i\i)$ a
di-neutron, and $(\i\i\i)$ a tri-neutron,
becomes increasingly attractive.
By deducing from the uncertainty in two-neutron $S$-wave scattering-length
extractions -- in lattice calculations and experiments -- an even less
accurately constrained $P$-wave interaction,
we find bound ${}^3n$'s and ${}^4n$'s sustainable within this uncertainty.
The formation of these states is independent of the existence
of a bound $^1S_0$ di-neutron\footnote{For this conclusion, we weakened the ${}^1S_0$
attraction of the heavy-pion EFT until ${}^2n$ became unbound, too.}, and
hence the scenario of a bound tetra-neutron with none of its subsystems bound 
realizes a Brunnian nuclear state.

It remains to translate this effect to physical systems.
In other words, we release the fixed-node condition, as imposed
through the stronger attraction between neutrons at a pion mass
of 806~MeV, and
calibrate the interaction strengths $c_{S,T}$ to the
deuteron and the virtual singlet state with $k\approx -i\,25~$MeV.
We constrain $\eps$ with experimental $P$-wave data~\cite{PhysRevC.93.045201}
as shown in Fig.~\ref{fig:2n-phases}.
In the subsequent analysis of the ${}^2n({}^3P_0)$,
the ${}^3n(\nicefrac{1}{2}^-)$, and the the ${}^4n(0^+)$ states, which
were identified as most sensitive\footnote{Ref.~\cite{PhysRev.150.839}
comes to the opposite conclusion although within a model-dependent approach.}
to the arbitrarily small
spin-orbit disturbance, we find the same signature of the
molecular character of the ${}^2n$ interaction as for
the heavy-quark system. Namely, critical short-distance
behavior appears in the three systems at well spaced values of $\eps$.
This means, \eg, that for $\eps\approx\eps({}^3n)$, a shallowly
bound tri-neutron can exist, but no di- or tetra-neutron. The latter
are either unbound or deeply\footnote{In principle, such a state could become
significant as yet another~\cite{Detmold:2014qqa,Appelquist:2015yfa}~dark-matter candidate
of composite
Standard-Model degrees of freedom.} bound. Furthermore, the hierarchy
\be\la{eq:eps-hie}
\eps({}^4n)<\eps({}^3n)<\eps({}^2n)\;\;,
\ee
which encodes the appearance of bound states in larger systems
for weaker two-neutron interactions, is found independent of
$\Lambda$, the parameter which complements $\eps$ in the
uncertainty assessment\footnote{$\Lambda\to\infty$ corresponds to the so-called zero-range approximation,
where universal behavior is identified (see the review~\cite{Braaten:2004rn}).}.
This independence implies that a weak attraction
between two neutrons becomes more attractive
between three and even stronger between four neutrons as a consequence
of the physical deuteron binding energy, the two-nucleon singlet scattering
length, and the specific nucleon mass. With the assumption that
the enhancement is proportional to the overlap of the
two-neutron state most sensitive to the weak attraction,
and thereby proportional to the number of two-neutron pairs in
this configuration -- above, we considered the ${}^3P_0$ state --
the emergence of a bound \cjRL{b*}-neutron state follows.

How does this prediction of a bound neutron nucleus relate to previous
theoretical work?
On the one hand, the most advanced -- in terms of the accuracy of the employed
interaction models and the precision of the numerical method -- theoretical
investigations~\cite{PhysRevLett.90.252501,PhysRevC.93.044004,Gandolfi:2016bth}~do not find neutron nuclei bound.
Consistently, calculations of
equations of state of neutron- and asymmetric nuclear matter with modern 
high-accuracy interaction models do not show
characteristics of emergent, bound neutron clusters
(compare, \eg, Refs.~\cite{Gandolfi:2013baa,Lynn:2015jua,PhysRevLett.110.032504,PhysRevLett.111.032501}~
and~\cite{RevModPhys.50.107,Ogloblin1989}).
On the other hand, Ref.~\cite{Ogloblin1989}~compiles earlier work~\cite{Baz:1972ag}~
based on the
already mentioned intuitive analogy between the formation of bound neutron drops
and liquid helium, which suggested relatively large bound neutron nuclei
with $\text{\cjRL{b*}}>64$.
In Ref.~\cite{PhysRevC.71.044004}, it was shown more recently how
a similar modification of the nuclear interaction, namely a
selective enhancement of a single $P$-wave channel, can lead to
bound tri-neutrons without severe consequences for two-nucleon observables.
From these analyses and our work, two classes of modifications of nuclear interaction
models can be identified. Modifications of the \emph{first} class bind neutron nuclei but affect
ordinary nuclei. The resultant unphysical models modify either the ${}^1S_0$ component
of the NN interaction, enhance the $P$-wave NN channels democratically,
or they introduce $T=\nicefrac{3}{2}$ three-nucleon or $T=2$ four-nucleon interactions.
Modifications of the \emph{second} class, selectively enhance the interaction in two-neutron
channels with an odd relative angular momentum subject to constraints by data on
ordinary nuclei. The specific operator which induces this behavior is irrelevant.
In the EFT framework, any operator which modifies the short-distance structure of the
observables of interest consistently with the accuracy of the considered order of
the EFT expansion is admissible.
A future realization of a second class modification with a high-accuracy
model can resolve the seeming contradiction in the theoretical analyses. 

Relating to experiment, we stress that neither this nor all previous work
predict the precise number of neutrons necessary to bind a neutron nucleus.
It is the sheer existence which is predicted.
While a direct detection of such a potentially
large droplet might be as hard as resolving the conflicting measurements on tri- and
tetra-neutrons or resonances\footnote{Experimental findings of such shallow structures are
reported in
Refs.~\cite{PRC.65.044006,Novatsky2012,Novatsky2014,BYSTRITSKY2016164,PhysRevLett.116.052501}
in contrast to negative measurements in
Refs.~\cite{Yuly:1997ja,Grater:1999cz,Grater:1999xi,Aleksandrov2005}~(We
list only work known to us and not reviewed in Refs.~\cite{Tilley:1987svb,Tilley:1992zz}).},
the inference of the peculiar Brunnian character can be tested in fermionic systems
in which a modification of the interaction of the described type is feasible. Loss rates associated with
the formation of bound multi-fermion cluster\footnote{Consistent
with Ref.~\cite{PhysRevA.86.012710}, the $\eps$ tuning does not yield an accumulation of
shallow ${}^3n$ states.} for a finely tuned\footnote{Exemplary were
experiments on cold-atom gases~\cite{PhysRevLett.102.140401},
where a magnetic field was used to
tune the atom-atom interaction to detect the Efimov effect.} two-fermion $P$-wave
interaction are signatures of the effect.

The critical number \cjRL{b*} can be predicted with available data which is
correlated to the two-fermion $P$-waves but more accurately measured.
Nuclear $P$-wave observables,
like the first excited state of helium-4, the resonant states
of helium-5, and/or the instability of beryllium-8 suggest themselves.
Assuming the NN $P$-waves to be constrained by any of these systems, future work
will enable a prediction of the minimal neutron number \cjRL{b*}.

Before summarizing, it is in order to list to important assumptions.
$(\i)$ Isospin-breaking effects must lead to similar or larger deviations 
of the ${}^2n$ $P$-wave phases from the respective neutron-proton phase shifts.
Evidence for this is provided by the relatively small shift of the summed phases
necessary to yield bound multi-neutron. $(\i\i)$ The
increasing probability to find neutron
pairs\footnote{The success of such an approach in the description
of bulk features of the nuclear chart has been demonstrated in Ref.~\cite{Wiringa:2006ih}.}
in relative $P$-waves in larger systems and the change induced in the two-neutron
interaction by the background of the other neutrons must eventually
accumulate to an interaction corresponding to a critical $\eps$.
We have shown this explicitly for the most crucial numbers, from the shell-model
perspective, three and four.
Under these two assumptions,
we predict a multi-neutron state with binding energy of order 10~MeV
and thereby of a scale comparable to
the deuteron and triton ground states,
and the bonding energy per nucleon of
larger nuclei in a narrow window around the respective
critical $\eps({}^\text{\cjRL{b*}}n)$.
For other $\eps$, despite practically invariant NN phase shifts,
the shallow neutron clusters become very deeply bound
and thus decouple from processes at the nuclear scale
and might also escape a detection in numerical calculations.
\parag{Epilogue}\la{sec:sum}
We fine tune the nuclear interaction such that it supports bound multi-neutron
states with energies of order 10-100~MeV without sacrificing the usefulness
of the theory for its description of ordinary nuclei. The binding mechanism is attributed
to a specific component of the two-fermion interaction that becomes increasingly attractive with the
number of particles in the system. For the three and four-neutron system, we find the enhancement factors
sufficiently different to infer that only one of the two can be shallowly bound,
while the other is either unbound or so deeply bound that its coupling to nuclear processes can
be disregarded. We extrapolate this finding to larger systems and deduce the existence
of a finite number of neutrons which is bound, while any of its isolated sub-configurations
is unbound. The employed interaction model suggests the emergence of such
Brunnian states in any two-component fermionic system with an entirely attractive two-body
potential with a characteristically large scale relative to its range.
\section*{Acknowledgements}
I owe thanks to B.~C.~Tiburzi for his careful reading of and comments on the manuscript.
Support comes from the National Science Foundation under Grant No. PHY15-15738.
\section{Appendix}
\parag{Interaction}
In table~\ref{tab:LECs}, we list the low-energy constants (LEC) used in our calculations.
The triplet constant
$c^\Lambda_{T}$ is calibrated to the deuteron binding energy, $\bd=2.22~$MeV in the
physical world, and $\bd=19.5~$MeV in the heavy-pion world. The singlet LEC $c^\Lambda_{S}$
is fitted to the $S$-wave neutron-proton singlet scattering length $\at=-23.8~\text{fm}$ for
physical quarks and to the binding energy of the singlet neutron-proton ground state at
\mpil, namely $B_S=16~$MeV.
The three-nucleon contact operator 
\be\la{eq:tni}
\hat{V}_3= \sum_{i<j<k}^A \sum_\text{cyc}
d_3(\Lambda)\,\big(\nicefrac{1}{2}-\nicefrac{1}{6}\,\ve{\tau}_i\cdot\ve{\tau}_j\big)
e^{-\frac{\Lambda^2}{4}\left(\ve{r}_{ij}^2+\ve{r}_{ik}^2\right)}\;,
\ee
projected onto the spin-doublet channel, is included in the calculations
to assess the invariance of the triton and helium-4 binding energies with respect to the
disturbance induced through the spin-orbit interaction.
Its strength $d(\Lambda)_{3}$ is fitted to match
the triton's binding energy $\bt=8.48~$MeV at the physical and $\bt=53.9~$MeV
at \mpil.

The superscript $\Lambda$ parameterizes a Gaussian regulator
of the singular contact interaction and a substitution
\be\la{eq:regul}
c^\Lambda\to c(\Lambda)\,e^{-\frac{\Lambda^2}{4}\ve{r}^2}
\ee
is understood with the relative distance $\ve{r}$ between two interacting particles.
We employ interactions with $\Lambda\in[6,12]~\text{fm}^{-1}$ because
lower cutoffs were shown to unjustly cut out modes at large pion masses
which are necessary for the formation of the
different bound states. To observe a convergent behavior it is unnecessary to go beyond
$12~\text{fm}^{-1}$ where numerical convergence is harder to demonstrate than
for values closer to the physical scales.
\begin{table}[h]
\begin{center}
\caption{The 
LECs $c_{S,T}(\Lambda)$, $d_3(\Lambda)$ [GeV] 
for physical ($m_{\pi}=140$ MeV) and lattice
($m_{\pi}=806\;\rm{MeV}$) nuclei 
for various values 
of the momentum cutoff 
\mbox{$\Lambda$ [fm$^{-1}$]}.}
\label{tab:LECs}
\begin{tabular}
{c@{\hspace{5mm}} c@{\hspace{5mm}} c@{\hspace{5mm}} c@{\hspace{5mm}} c@{\hspace{5mm}}}\hline\hline
$m_{\pi}$ & $\Lambda$ &  $c_T(\Lambda)$  &  $c_S{\Lambda}$ & $d_3{\Lambda}$ \\
\hline
  140   &$2$ &$-0.1423$ &$-0.1063$ &$0.06849$\\ 
        &$4$ &$-0.5051$ &$-0.4350$ &$0.6778$ \\ 
        &$6$ &$-1.091$ &$-0.9863$ &$2.653$   \\
        &$8$ &$-1.899$ &$-1.760$ &$7.816$    \\
        &$10$ &$-2.929$ &$-2.757$ &$20.48$   \\
        &$12$ &$-4.182$ &$-3.976$ &$50.94$   \\
        &$15$ &$-6.480$ &$-6.222$ &$195.6$   \\
\hline
  806   &$2$  &$-0.1480$ &$-0.1382$ &$ 0.07102$\\
        &$4$  &$-0.4046$ &$-0.3885$ &$ 0.3539$ \\
        &$6$  &$-0.7892$ &$-0.7668$ &$ 1.001$  \\
        &$8$  &$-1.302$  &$-1.273$  &$ 2.221$  \\
        &$10$ &$-1.942$  &$-1.907$  &$ 4.308$  \\
        &$12$ &$-2.710$  &$-2.670$  &$ 7.712$  \\
        &$15$ &$-4.103$  &$-4.052$  &$16.84$   \\
\hline\hline
\end{tabular}
\end{center}
\end{table}

\parag{Resonating groups}
A version~\cite{Hofmann:1986} of Wheeler's resonating-group method~\cite{Wheeler:1937zz}
is used to solve the stationary
Schr\"odinger equation.
For the considered systems, the general
two-fragment ansatz for the wave function,
\be\la{eq:rgwf}
\Psi=\mathcal{A}\left\lbrace\sum_i\phi_I^{(i)}\phi_{II}^{(i)}F^{(i)}(\ve{R}_i)\right\rbrace\;\;,
\ee
simplifies because the second fragment $\phi_{II}$ is identified with
a single neutron\footnote{We find ${}^2n-{}^2n$ configurations insignificant in the
considered ${}^4n$ channels.}.
The first fragment $\phi_I$ is either another neutron, a ${}^1S_0$ dineutron, or a
$\nicefrac{1}{2}^-$ trineutron. For the latter, $i$ indexes all spin coupling schemes of
three spin$-\nicefrac{1}{2}$ neutrons which couple with a
total angular momentum $L=1$ to $J=\nicefrac{1}{2}$.
In the partial-wave expansion of the inter-cluster relative-motion function
\be\la{eq:relm}
F(\ve{R})=\frac{1}{R}\sum_lf_l(R)P_l(\cos\theta_R)
\ee
we find sufficiently converged results with $l\leq1$.
In addition to these functions,
which describe asymptotically non-interacting fragments,
\mbox{$\hat{H}\Psi=\left[E_\text{\tiny c.m.}-B({}^{2(3)}n)\right]\Psi$},
so-called distortion channels are included in the sum over $i$. For those, the
functions $F^{(i)}$ do not approximate a free relative motion, but allow for
any deviation from the ``frozen'' configuration when fragments are close and
interact strongly. Each fragment is set up in an antisymmetric state, and the
cross-fragment antisymmetrizer $\mathcal{A}$ excludes, \eg, all dineutron-dineutron
configurations with positive parity from the ${}^4n$ $0^+$ channel.
We expand the radial functions $f_l$, which appear also in the cluster-internal
functions $\phi$, in a Gaussian basis whose parameters are optimized for each cutoff
parameter $\Lambda$. A typical dimension of a variational space in which results
converge is found to be of $\mathcal{O}(10^3)$ because of this tailoring.
For the scattering problem, a variant of the Kohn-Hulth\'{e}n variational principle
is invoked, and bound-state energies are inferred from a diagonalization of $\hat{H}$
in the basis comprised of the element space of $i$ in Eq.\eqref{eq:rgwf} (spin/orbital
angular-momentum coupling schemes, Gaussian parameters).
%
%
\bibliographystyle{apsrev4-1}
\bibliography{refs}
\end{document}